\documentclass{emulateapj}
\usepackage{graphicx, color}
\usepackage{longtable}
\usepackage{amssymb}
\usepackage{tabularx, blindtext}

\begin{document}
\shorttitle{SL analysis of MACS J0416.1-2403}
\shortauthors{Zitrin et al.}

\slugcomment{Submitted to the Astrophysical Journal Letters}

\title{CLASH: The enhanced lensing efficiency of the highly elongated merging cluster \\ MACS J0416.1-2403}

%\author{A. Zitrin\altaffilmark{1}}
%\author{J. Moustakas\altaffilmark{2}}
%\author{L. Bradley\altaffilmark{3}}
%\author{D. Coe\altaffilmark{3}}
%\author{L.A. Moustakas\altaffilmark{4}}
%\author{M. Postman\altaffilmark{3}}
%\author{X. Shu\altaffilmark{5}}

\author{A. Zitrin\altaffilmark{1,*}, M. Meneghetti\altaffilmark{2}, K. Umetsu\altaffilmark{3}, T. Broadhurst\altaffilmark{4,5}, M. Bartelmann\altaffilmark{1}, R. Bouwens\altaffilmark{6}, L. Bradley\altaffilmark{7}, M. Carrasco\altaffilmark{1,8}, D. Coe\altaffilmark{7}, H. Ford\altaffilmark{9}, D. Kelson\altaffilmark{10}, A.M. Koekemoer\altaffilmark{7}, E. Medezinski\altaffilmark{9}, J. Moustakas\altaffilmark{11}, L.A. Moustakas\altaffilmark{12}, M. Nonino\altaffilmark{13}, M. Postman\altaffilmark{7}, P. Rosati\altaffilmark{14}, G. Seidel\altaffilmark{15}, S. Seitz\altaffilmark{16,17}, I. Sendra\altaffilmark{4}, X. Shu\altaffilmark{18}, J. Vega\altaffilmark{19,20}, W. Zheng\altaffilmark{9}}

%\affil{
\altaffiltext{1}{Universit\"at Heidelberg, Institut f\"ur Theoretische Astrophysik, Heidelberg, Germany; adizitrin@gmail.com}
\altaffiltext{2}{INAF, Osservatorio Astronomico di Bologna, \& INFN, Sezione di Bologna; Bologna, Italy}
\altaffiltext{3}{Institute of Astronomy and Astrophysics, Academia Sinica, Taipei, Taiwan}
\altaffiltext{4}{Department of Theoretical Physics, University of Basque Country UPV/EHU, Bilbao, Spain}
\altaffiltext{5}{IKERBASQUE, Basque Foundation for Science, Bilbao, Spain}
\altaffiltext{6}{Leiden Observatory, Leiden University, Leiden, Netherlands}
\altaffiltext{7}{Space Telescope Science Institute, Baltimore, MD, USA}
\altaffiltext{8}{Department of Astronomy and Astrophysics, AIUC, Pontificia Universidad Cat\'olica de Chile, Santiago, Chile}
\altaffiltext{9}{Department of Physics and Astronomy, The Johns Hopkins University, Baltimore, MD, USA}
\altaffiltext{10}{Observatories of the Carnegie Institution of Washington, Pasadena, CA, USA}
\altaffiltext{11}{Department of Physics \& Astronomy, Siena College, Loudonville, NY, USA}
\altaffiltext{12}{Jet Propulsion Laboratory, California Institute of Technology, Pasadena, CA, USA}
\altaffiltext{13}{INAF-Osservatorio Astronomico di Trieste, Trieste, Italy}
\altaffiltext{14}{ESO-European Southern Observatory, Garching bei M\"unchen, Germany}
\altaffiltext{15}{Max-Planck-Institute for Astronomy, Heidelberg, Germany}
\altaffiltext{16}{University Observatory Munich, M\"unchen, Germany}
\altaffiltext{17}{Max-Planck-Institut f\"ur extraterrestrische Physik, Garching, Germany}
\altaffiltext{18}{Department of Astronomy, University of Science and Technology of China, Hefei, Anhui, China}
\altaffiltext{19}{Departamento de F\'isica Te\'orica, Universidad Aut\'onoma de Madrid, Madrid, Spain}
\altaffiltext{20}{LERMA, Observatoire de Paris, Paris, France}
\altaffiltext{*}{\textbf{Mass models are publicly available at: ftp://wise-ftp.tau.ac.il/pub/adiz/M0416/}}

%\footnote{An example**}
\begin{abstract}
We perform a strong-lensing analysis of the merging galaxy cluster MACS J0416.1-2403 (M0416; $z=0.42$) in recent CLASH/HST observations. We identify 70 new multiple images and candidates of 23 background sources in the range $0.7\lesssim z_{phot} \lesssim 6.14$ including two probable high-redshift dropouts, revealing a highly elongated lens with axis ratio $\simeq$5:1, and a major axis of $\sim100\arcsec$ ($z_{s}\sim2$). Compared to other well-studied clusters, M0416 shows an enhanced lensing efficiency. Although the critical area is not particularly large ($\simeq0.6~\square\arcmin$; $z_{s}\sim2$), the number of multiple images, per critical area, is anomalously high. We calculate that the observed elongation boosts the number of multiple images, \emph{per critical area}, by a factor of $\sim2.5\times$, due to the increased ratio of the caustic area relative to the critical area. Additionally, we find that the observed separation between the two main mass components enlarges the critical area by a factor of $\sim2$. These geometrical effects can account for the high number (density) of multiple images observed. We find in numerical simulations, that only $\sim4\%$ of the clusters (with $M_{vir}\geq6 \times10^{14} h^{-1}M_{\odot}$) exhibit as elongated critical curves as M0416.
\end{abstract}

%\begin{keywords}
%\end{keywords}

%% text by Moustakas
%\def\imtxt#1{{\bf {\textcolor{red}{#1}}}}
%\def\imsout#1{{\bf {\textcolor{red}{\sout{#1}}}}}

\section{Introduction}\label{intro}

Due to their high, projected surface mass densities, galaxy clusters magnify and distort background objects, forming natural gravitational lenses in the sky. The lensing and magnification effects generally increase towards the central region of the cluster, where the projected mass density is often high enough to form multiple images of the same background source, depending also on the angular diameter distances involved (for reviews see \citealt{Bartelmann2010reviewB}; \citealt{Kneib2011review}).

The lensing efficiency of galaxy clusters (e.g. the number of multiple-images generated) is also related to other factors, such as the ellipticity, amount of substructure and its distance from the center, and degree of relaxation or merger \citep[e.g.][]{Meneghetti2003}. For example, the critical area grows with the concentration of the cluster \citep[e.g.][]{SadehRephaeli2008}, so it is clear that massive and more concentrated clusters should show more multiple images. The ``over-concentration'' problem in which lensing-selected clusters are found to have high concentrations \citep[and large Einstein radii, e.g.][]{ComerfordNatarajan2007CMrelation,Broadhurst2008, BroadhurstBarkana2008}, is often attributed to a lensing selection bias towards higher concentrations of triaxial clusters preferentially aligned with the line of sight \citep[see also][and references therein]{Hennawi2007,Sereno2010,Oguri201238clusters,Okabe2010}.

On the other hand, recent efforts show that there exists a second class of prominent strong lenses. \citet{SerenoZitrin2012} showed that in a triaxial lensing analysis, the 12 MACS clusters at $z>0.5$ \citep[][]{EbelingMacs12_2007} have relatively low concentrations, despite the many multiple images uncovered in their fields. They suggested that since most of these clusters are not yet relaxed, the amount of substructure in their centers as well as their higher redshift than most previously known lenses, turn them into highly-magnifying lenses. The critical curves of the several subclumps are merged together into a bigger lens, whose overall mass profile in the central part is often shallower thus boosting the magnification.

Indeed, various clusters around $z\sim0.5$ show prominent lensing features \citep[giant arcs and many multiple images;][]{Zitrin2012CLASH0329,Zitrin2012CLASH1206}. Interestingly, the largest gravitational lens, MACS J0717.5+3745 \citep[$z=0.55$;][]{Zitrin2009_macs0717}, is formed by several merging clumps possibly at the tip of a filament \citep{Limousin2012_M0717,Jauzac2012WL0717}. \citet{Zitrin2009_macs11495} showed that MACS J1149.5+2223 ($z=0.54$) is an excellent ``magnifying glass'' in the sky due to its shallow inner mass profile. This cluster is now known to magnify a galaxy at $z\sim10$ (\citealt{Zheng2012NaturZ}). Recently, \citet{Coe2012highz} uncovered the highest redshift galaxy known to date at $z\sim10.8$, multiply-imaged by MACS J0647.7+7015 ($z=0.59$), another complex cluster expected to be highly magnifying \citep{Zitrin2011_12macsclusters,PostmanCLASHoverview}.

\citet{Meneghetti2003} found that the cluster lensing cross section for giant arcs grows rapidly with ellipticity. A typical ellipticity of $e=1-b/a=0.3$ entails an order of magnitude increase in the lensing cross section \citep[see also][]{Torri2004}, for example, because with increasing ellipticity the caustics stretch, develop cusps, and enclose a growing area. \citet{Meneghetti2003} also found, that approaching subclumps boost the lensing cross sections. \cite{meneghetti2007} examined the arc sensitivity to cluster ellipticity, asymmetries, and substructures, and found that these contribute, respectively, $\sim40\%$, $\sim10\%$, and $\sim30\%$, to the lensing cross section. Recently, \citet{Redlich2012MergerRE} found that cluster mergers can enhance the lensing cross section by typically $\sim30-50\%$. It is therefore expected that merging, substructured and elongated clusters, should also constitute prominent strong lenses.

Here, we present the lensing analysis (\S 2) of the merging cluster MACS J0416.1-2403 (hereafter M0416; \citealt{MannEbeling2012evolution}), performed on recent Hubble Space Telescope (HST) imaging in 16 bands from the UV to the near-IR to a total depth of $\sim20$ orbits, as part of the Cluster Lensing And Supernova survey with Hubble (CLASH) program \citep[see][]{PostmanCLASHoverview}. The CLASH pipeline uses the 16-band observations to derive photometric redshifts for each arc, via the BPZ program \citep{Benitez2004,Coe2006}, used here to constrain the model. As we shall see, the cluster exhibits high elongation, in part as a result of the possible merger. We test, by a simple semi-analytical simulation (\S 3), the expected increase in lensing efficiency with ellipticity and merger stage, to see if these may account for the many multiple images in M0416, per its critical area. For this work, we define the \emph{lensing efficiency} simply as \emph{the number of multiple images per critical area}, which we also refer to as \emph{the number density} of multiple images. This ratio, naturally, scales with the ratio of the caustic area and the critical area.

M0416 was listed as a MACS cluster (MAssive Cluster Survey, see \citealt[][]{Ebeling2010FinalMACS}) due to its X-ray brightness. \citet{MannEbeling2012evolution} classified it as merging based on the double-peaked X-ray structure, where the southern peak is offset by a few arcseconds from the corresponding (second) BCG, as expected in mergers \citep[e.g.][]{Bradac2006Bullet,Merten2011}. Based also on its predicted Einstein radius, M0416 was designated as one of the five ``high-magnification'' CLASH clusters (where 20 clusters are X-ray selected to be dynamically relaxed). We found no record of a published strong lensing (SL) analysis of M0416, but \citet{Christensen2012Specs} published a measured spectroscopic redshift for the giant arc north of the BCG (systems 1 \& 2 here), $z_{s}=1.896$, which we use in our analysis.

\begin{figure*}
 \begin{center}
   \includegraphics[width=160mm]{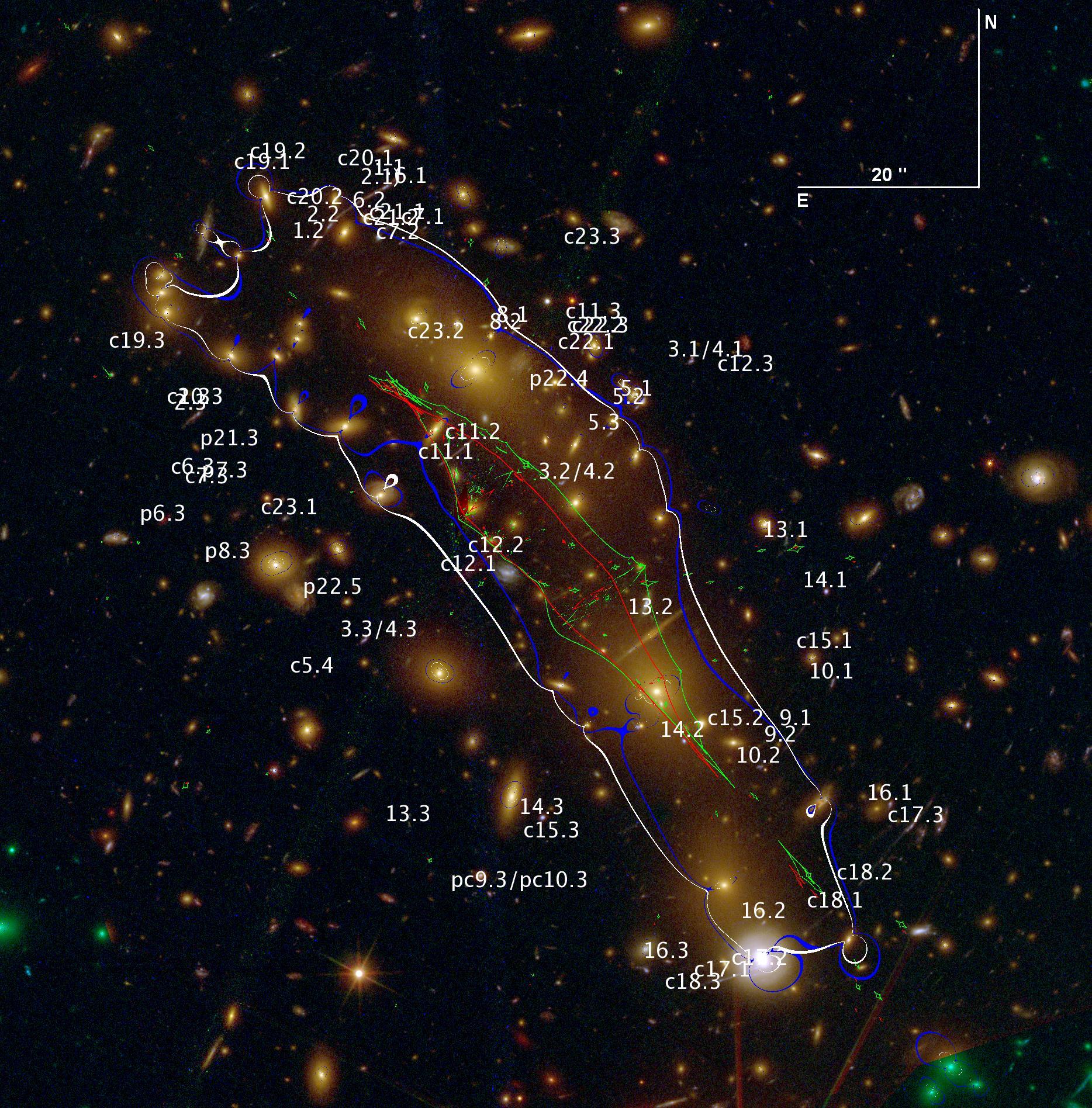}
 \end{center}
\caption{CLASH/HST 16-band color-composite image of M0416, with multiple images numbered. Lensed candidates are marked with ``c'', and ``p'' stands for predicted location. Overlaid in \emph{blue(white)} is the critical curve for $z_{s}=1.896$, corresponding to the giant arc [system 1\&2], from the eNFW(eGaussian) based model (see \S \ref{SLanalysis}). In \emph{red(green)} we plot the corresponding caustics.}
\label{multipleimages}
\end{figure*}

Throughout we adopt a concordance $\Lambda$CDM cosmology with ($\Omega_{\rm m0}=0.3$, $\Omega_{\Lambda 0}=0.7$, $h=0.7$), where $1\arcsec= 5.53$ kpc at the adopted redshift for M0416, $z=0.42$ \citep{Christensen2012Specs}.

\section{Strong Lensing Analysis of M0416}\label{SLanalysis}

We use two complementary modeling techniques to construct mass models for M0416. These are then compared to each other and to our new weak lensing (WL) analysis.

\emph{Method 1:} The first method we use follows the prescription of \citet{Zitrin2009_cl0024}, with several modifications recently implemented for speed and a wider choice of priors. First, instead of power-law profiles traditionally used to represent the galaxies in our method \citep{Broadhurst2005a,Zitrin2009_cl0024}, we model each galaxy with a Pseudo-Isothermal Elliptical Mass Distribution (PIEMD; \citealt{KassiolaKovner1993}), adopting the exact formulation from \citet[][]{Jullo2007Lenstool}:

\begin{equation}
\label{julloEqs}
\left\{ \begin{array}{l}
\sigma_0  =   \sigma_0^\star (\frac{L}{L^\star} )^{1/4}\;,  \\
r_{core}  =  r_{core}^\star (\frac{L}{L^\star} )^{1/2}\;, \\
r_{cut}  =   r_{cut}^\star (\frac{L}{L^\star} )^\alpha\;,  \\
\end{array}
\right.
\end{equation}
where $r_{core}$ is the core radius, $r_{cut}$ the cut-off radius, and $\sigma_0$ the velocity
 dispersion. The total mass of a subhalo then scales as:

\begin{equation}\label{julloEqs2}
 M = (\pi/G)(\sigma_0^\star)^2 r_{cut}^\star (L/L^\star)^{1/2+\alpha}\;,
\end{equation}
where $L^\star$ is the typical luminosity of a galaxy at
the cluster redshift, and $r_{cut}^\star$, $r_{core}^\star$ and
$\sigma_0^\star$ are its PIEMD parameters \citep{Jullo2007Lenstool}.

The PIEMD representation is used in many modeling methods \citep[e.g.][]{Halkola2006,Jullo2007Lenstool,Richard2010A370,Oguri2012SL}, and eases the comparison to the second method we implement here. As a second change, this mass distribution is now smoothed with an \emph{elliptical Gaussian} (eGaussian), instead of our traditional smoothing by a 2D polynomial spline, and therefore eliminates the need for an external shear as the ellipticity is now directly introduced into the mass distribution. The superposed galaxies' contribution and its eGaussian-smoothed map - are then added with a relative weight that is a free parameter. This method therefore includes 6 free basic parameters: $r_{cut}^{*}$, $\sigma_{0}^{*}$ for the PIEMD galaxy models (eqs. \ref{julloEqs} and \ref{julloEqs2}); $\sigma_{MJA}$ and $\sigma_{MNA}$, the widths of the eGaussian kernel along the major and minor axes; $\phi$, the angle of the major axis in the eGaussian kernel; and the relative fraction of the galaxies component from the total mass (the remaining fraction is contributed by the smooth DM component).

\begin{figure}
 \begin{center}
   \includegraphics[width=87mm]{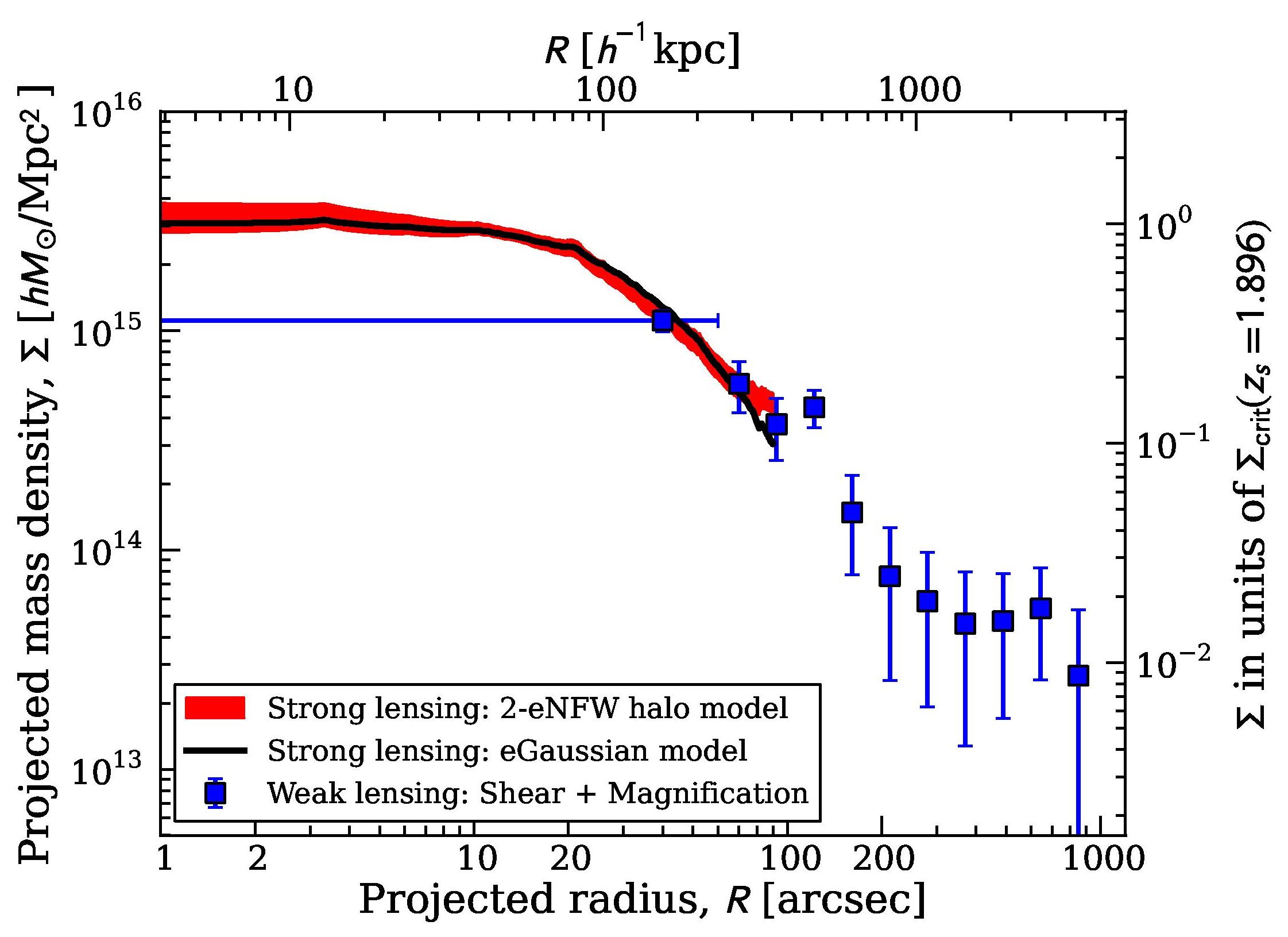}
 \end{center}
\caption{Projected mass density profile of M0416. The \emph{red} curve shows the resulting profile and 1$\sigma$ errors from the two eNFW halos model; and the \emph{black} curve shows the resulting profile from the eGaussian-smoothing model (\S \ref{SLanalysis} for details). The two models are similar in the range where multiple-image constraints are available ($<1\arcmin$), and highly consistent with independent Subaru WL analysis data (\emph{blue} squares and errorbars).}
\label{profile}
\end{figure}

\emph{Method 2:} The second method we use adopts the following parametrization. Galaxies are modeled each as PIEMD scaled by its light as above (Method 1; eq. \ref{julloEqs}). To represent the DM we implement an elliptical NFW (eNFW) halo. Since M0416 is clearly a complex, likely merging system, the first few modeling attempts introduced the need to add a second eNFW halo to the model. In total, then, the model includes the galaxy component described by the superposition of all PIEMD representations, and two eNFW halos (where the ellipticity, defined here throughout as $e=(MJA-MNA)/(MJA+MNA)$, is directly introduced into the NFW mass distribution via the transformation $r \rightarrow r_e=\sqrt{[x/(1+e)]^2 +[y/(1-e)]^2}$). We maintain the eNFW halos centered on the first and second brightest cluster members, respectively. This keeps the number of free parameters as low as possible and yields an excellent fit. Note that we tried to construct a model while allowing the (southern halo) center to vary but did not find a significantly better solution. This parametrization therefore consists of 10 free basic parameters: $r_{cut}^{*}$, $\sigma_{0}^{*}$ for the PIEMD galaxy models; the scale radius $r_{s}$ and the concentration parameter $c_{vir}$, as well as the ellipticity and its position angle, for each of the two eNFW halos.

Using a preliminary model with the \citet{Zitrin2009_cl0024} method, and the two iteratively-improved models above, along with a complementary examination by eye, we iteratively matched together multiply-imaged systems. The best fit solution in both methods is obtained via a long (several dozens of thousands steps) Monte-Carlo Markov Chain (MCMC) minimization, using several chains. We note that in some chains, including the final ones used here, the redshifts of some of the multiple-systems were left as free parameters (with flat priors) to be optimized. Also, in both final chains for methods 1 and 2, we left the three BCGs to be freely weighted and optimized by the MCMC. For the two main BCGs, we fix the ellipticity and position angle to the parameters derived using SExtractor \citep{BertinArnouts1996Sextractor}. Throughout, we fix for the PIEMD, $r_{core}^{*}=0.3$ kpc, use an $L^{*}$ value equivalent to an absolute magnitude of $M^{*}_{F814W}=-20.113$, and assume a constant mass-to-light ratio ($\alpha=0.5$, eq. \ref{julloEqs2}). The F814W-F475W color was used to extract the red-sequence, where we used here the 122 brightest (F814W) members.
\begin{table*}\tiny
\caption{Multiple Images and Candidates found by our method.}\tiny
\label{multTable}
\centering
\begin{tabularx}{0.95 \linewidth}{|c|c|c|c|c|c|c|c|l|}
\hline
Arc ID & RA(J2000.0) & DEC(J2000.0)  & phot-$z$ [$z_{min}$-$z_{max}$] & $z_{NFW}$ & $\Delta$ NFW $\arcsec$ & $z_{Gauss}$ & $\Delta$ Gauss $\arcsec$ & comments\\
\hline\hline
1.1 & 04:16:09.780 & -24:03:41.73 & 1.788 [1.541--1.879] &  2.01 [1.66--2.69] & 1.6 & 2.10 [1.67--2.17] & 1.5 & sys fixed to $z_{spec}$=1.896\\
1.2 & 04:16:10.435 & -24:03:48.75 & 2.482 [2.379--2.560] & " & 0.9& " & 2.3&nearby part $z_{phot}\sim1.9$\\
1.3 & 04:16:11.365 & -24:04:07.21 & 2.555 [2.442--2.675] & " & 0.3 & "& 0.9&nearby part $z_{phot}\sim1.9$\\
2.1 & 04:16:09.884 & -24:03:42.77 & 1.788 [1.541--1.879]  & 2.00 [1.71--2.13] &1 & 2.24 [1.73--2.24] &0.5 &\\
2.2 & 04:16:10.321 & -24:03:46.93 & 1.846 [1.796--1.998]  & "& 0.5& "&1.2 &\\
2.3 & 04:16:11.394 & -24:04:07.86 & 1.928 [1.806--2.239]  & " & 0.7&" &1.4 &\\
\hline
3.1 & 04:16:07.388 & -24:04:01.62 & 2.149 [2.130--2.323] &  2.00 [1.61--3.74] &  1.1& 2.14 [1.59--3.97]& 0.9 &\\
3.2 & 04:16:08.461 & -24:04:15.53 & 2.324 [2.166--2.369] & "& 0.5& "&2.3&blended with 4.2\\
3.3 & 04:16:10.036 & -24:04:32.56 & 2.778 [2.759--2.814] & "& 0.6& "&3&\\
4.1 & 04:16:07.398 & -24:04:02.01 & 2.199 [1.182--2.819] & 1.99 [1.60--3.54] & 0.4& 1.81 [1.58--4.03]&2.3&\\
4.2 & 04:16:08.437 & -24:04:15.53 & 2.324 [2.166--2.369] & "& 1& "&0.7&blended with 3.2\\
4.3 & 04:16:10.051 & -24:04:33.08 & 2.244 [2.140--2.325] & "& 0.5& "&0.9&\\
\hline
5.1 & 04:16:07.773 & -24:04:06.24 & --- &  2.32 [1.74--3.63] & 1.4& 2.54 [2.07--3.53] &0.5&blended with member\\
5.2 & 04:16:07.839 & -24:04:07.21 & 2.362 [2.158--2.489] & "& 2.5& "&2.6&\\
5.3 & 04:16:08.043 & -24:04:10.01 & 2.565 [2.469--2.733] & "& 0.6& "&0.5&\\
c5.4 & 04:16:10.454 & -24:04:37.05 & 2.166 [1.860--2.601] & "& 0.7 &"&0.6&very plausible\\
\hline
6.1 & 04:16:09.609 & -24:03:42.64 & 6.139 [0.804--6.662] & 6.76 [3.81--7.31] & 2.0& 6.29 [3.99--17.30]& 2.5&bimodal $z_{phot}$\\
6.2 & 04:16:09.946 & -24:03:45.31 & --- & "& 0.7& "& 0.4&\\
p6.2 & 04:16:11.698 & -24:04:21.12 & --- & "&--- & "&--- &\\
c6.3 & 04:16:11.422 & -24:04:14.95 & 1.470 [0.501--6.472] &&& "&& bimodal $z_{phot}$\\
\hline
c7.1 & 04:16:09.552 & -24:03:47.13 & 2.376 [2.079--2.698] &3.81 [3.36--4.63] &3.2& 4.18 [3.45--4.63]&2.6&\\
c7.2 & 04:16:09.752 & -24:03:48.82 & 1.825 [1.801--1.982] & "& 0.3&"& 0.7&\\
p7.3 & 04:16:11.166 & -24:04:15.40 & --- & "&--- &"&---&\\
c7.3 & 04:16:11.308 & -24:04:15.99 & 2.378 [0.792--2.994] & "& &"& &\\
\hline
8.1 & 04:16:08.783 & -24:03:58.05 & 2.425 [2.266--2.622] & 3.08 [1.32--3.41] &0.5 & 2.97 [2.51--3.41] & 1.3&\\
8.2 & 04:16:08.840 & -24:03:58.83 & 2.414 [2.083--2.593] & "& 1.1& "&0.3&\\
p8.3 & 04:16:11.266 & -24:04:24.70 & --- & "&--- & "&---&\\
\hline
9.1 & 04:16:06.486 & -24:04:42.90 & 2.304 [1.922--2.547] &  3.65 [2.87--4.33]& 0.8& 3.93 [3.28--4.33]&1.6&\\
9.2 & 04:16:06.605 & -24:04:44.78 & 2.386 [1.827--2.962] & "& 0.9& "&0.2&\\
p9.3 & 04:16:08.992 & -24:05:00.00 & ---&"& ---&"&---&\\
c9.3 & 04:16:09.149 & -24:05:01.23 & ---&"& &"&&other cand. nearby\\
\hline
10.1 & 04:16:06.244 & -24:04:37.76 & 2.357 [2.236--2.526] & 4.63 [1.98--4.80]&2.5& 4.63 [3.04--4.80]&1.4& mean chain $z$=3.13\\
10.2 & 04:16:06.833 & -24:04:47.12 & 2.356 [2.231--2.630] & "&0.9 &" &0.4&"\\
p10.3 & 04:16:09.158 & -24:05:00.45 & 2.356 [2.231--2.630] & "& ---&" &---&"\\
c10.3 & 04:16:08.807 & -24:05:01.94 & 2.124 [1.946--2.361] & "&& "& &\\
c10.3 & 04:16:09.818 & -24:04:58.69 & 2.345 [2.115--2.775] & "&& " &&other cand. nearby\\
\hline
c11.1 & 04:16:09.410 & -24:04:13.32 & 0.717 [0.666--0.784] & 1.04 [0.99--1.19] & 2.7&1.18 [1.17--1.20]& 1.8&\\
c11.2 & 04:16:09.196 & -24:04:11.11 & 1.008 [0.833--1.092] & "& 1.7&"& 1.0&\\
c11.3 & 04:16:08.214 & -24:03:57.66 & 0.619 [0.443--1.130] & "& 1.2& "&1.2&\\
\hline
12.1 & 04:16:09.230 & -24:04:25.74 & 1.618 [1.430--2.085] &  2.79 [1.94--3.41] & 0.6& 1.82 [1.49--1.87]&1.8&\\
12.2 & 04:16:09.011 & -24:04:23.72 & 1.765 [1.031--2.043] & "& 1.2& "&1.8&\\
p12.3 & 04:16:06.966 & -24:03:59.93 & 1.288 [0.432--3.334] & "& ---& "&---&\\
c12.3 & 04:16:06.989 & -24:04:03.57 &  & && "& &\\
\hline
13.1 & 04:16:06.619 & -24:04:22.03 & 3.255 [3.181--3.357] &  4.35 [2.76--5.55] & 1.2&3.24 [2.79--7.82]&1.1& mean chain $z$=3.83\\
13.2 & 04:16:07.711 & -24:04:30.61 & --- & "& 1.2& "&3.7&"\\
13.3 & 04:16:09.681 & -24:04:53.56 & 3.280 [3.008--3.481] & "&0.8& "&1.2&"\\
\hline
14.1 & 04:16:06.296 & -24:04:27.62 & 1.765 [1.728--1.786] &  1.94 [1.60--2.42] & 2.8&2.01 [1.81--2.27]& 1.1&\\
14.2 & 04:16:07.450 & -24:04:44.26 & 1.196 [1.093--1.243]& "& 2.8& "&1.3&bright member nearby\\
14.3 & 04:16:08.598 & -24:04:52.78 & 1.773 [1.738--1.790]& "& 1.2& "&0.9&\\
\hline
c15.1 & 04:16:06.292 & -24:04:33.67 & 0.525 [0.205--1.774] & 1.80 [1.39--1.95] & 2.1&1.83 [1.64--1.95]& 1.5&other cand. nearby\\
c15.2 & 04:16:07.065 & -24:04:42.90 & --- & "& 4.2&" &3.5&other cand. nearby\\
c15.3 & 04:16:08.560 & -24:04:55.38 & 2.905 [2.559--3.168] & "&3.0& "& 1.5&\\
\hline
16.1 & 04:16:05.774 & -24:04:51.22 & 1.964 [1.840--2.079] & 2.35 [1.77--3.16]&0.8& 2.97 [2.66--3.41]&1.9 &\\
16.2 & 04:16:06.799 & -24:05:04.35 & --- & "& 0.3& "&1.0&\\
16.3 & 04:16:07.583 & -24:05:08.77 & 1.803 [1.405--1.870] & "&0.5& "& 0.6&\\
\hline
c17.1 & 04:16:07.170 & -24:05:10.91 & --- & 2.84 [1.92--3.81] & 0.9& 3.87 [3.36--5.08]& 1.5&\\
c17.2 & 04:16:06.866 & -24:05:09.55 & ---  & "& 1.1&"& 1.9&\\
c17.3 & 04:16:05.599 & -24:04:53.69 & 2.467 [2.336--2.543] & "&2.5& "& 2.4& part yields $z_{phot}\sim3$\\
\hline
c18.1 & 04:16:06.258 & -24:05:03.24 & 2.824 [1.314--3.083] &  2.41 [1.61--4.47] & 0.4&3.01 [2.87--4.72] &0.8&\\
c18.2 & 04:16:06.016 & -24:05:00.06 & 2.766 [2.199--2.924] & "& 2.3& "&3.6&\\
c18.3 & 04:16:07.416 & -24:05:12.28 & --- & "& 0.6& "&---&\\
\hline
c19.1 & 04:16:10.909 & -24:03:41.08 & ---  &  3.01 [2.13--3.08] & 0.5&3.60 [2.32--3.65] &1.7&\\
c19.2 & 04:16:10.777 & -24:03:39.85 & 2.774 [2.291--2.963] & "&1.8& "& 3.8&\\
c19.3 & 04:16:11.925 & -24:04:00.91 & 2.129 [1.229--2.611] & "& 1.2& "&0.6&\\
\hline
c20.1 & 04:16:10.069 & -24:03:40.63 & --- & 2.56 [1.79--11.95] & 2.7& 2.37 [1.87--5.08]&3.8& mean chain $z$=5.00\\
c20.2 & 04:16:10.478 & -24:03:44.98 & 1.306 [1.006--2.128] & "&0.7& "& 0.9&\\
c20.3 & 04:16:11.451 & -24:04:07.15 & 0.815 [0.469--4.155] & "& 2.6&"& &bimodal $z_{phot}$\\
\hline
c21.1 & 04:16:09.813 & -24:03:46.67 & 2.787 [1.868--3.039] & 4.47 [1.54--5.19] & 1.3 & 3.45 [3.20--5.19]&1.1&mean chain $z$=3.91\\
c21.2 & 04:16:09.865 & -24:03:47.32 & 2.446 [2.066--2.670] & "& 0.3&"& 0.2&"\\
p21.3 & 04:16:11.185 & -24:04:11.82 & --- & "&--- &"&---&"\\
c21.3 & 04:16:11.047 & -24:04:07.73 & 3.061 [2.771--3.244] & "& & "&&\\
\hline
c22.1 & 04:16:08.278 & -24:04:01.07 & 0.362 [0.191--3.558] & 1.74 [1.53--1.79] & 3.8& 1.67 [1.52--1.64] &3.8&\\
c22.2 & 04:16:08.204 & -24:03:59.28 & --- & "& 2.3& "&2.3&\\
c22.3 & 04:16:08.162 & -24:03:59.22 & 3.172 [0.511--3.427] & "&2.6 &"& 2.6&\\
p22.4 & 04:16:08.508 & -24:04:05.00 & --- & "& ---& "&---&cand. nearby\\
p22.5 & 04:16:10.345 & -24:04:28.31 & --- & "& --- &"&---&\\
\hline
c23.1 & 04:16:10.691 & -24:04:19.56 & 2.410 [2.367--2.563] & 1.83 [1.23--2.45] &1.6 & 1.78 [1.16--2.54] & 1.0&\\
c23.2 & 04:16:09.505 & -24:03:59.87 & 1.143 [1.109--1.214] &" & 0.8 &"&0.7&other cand. nearby\\
c23.3 & 04:16:08.242 & -24:03:49.47 & 2.057 [1.884--2.427] & "& 1.1 &"& 1.8 &other cand. nearby\\
\hline\hline
\end{tabularx}\footnotesize
\tablecomments{$\emph{Column 1:}$ arc ID. ``c'' stands for candidate and ``p'' for predicted location. For \emph{candidates} the photo-$z$ distribution, or identification \emph{by eye}, was ambiguous; $\emph{Columns 2 \& 3:}$ RA and DEC in J2000.0; $\emph{Column 4:}$ photometric redshift and 95\% C.L.; $\emph{Column 5:}$ predicted and 95\% C.L. redshift by the eNFW model ; $\emph{Column 6:}$ reproduction distance of image from the observed location in the eNFW model; $\emph{Column 7:}$ predicted and 95\% C.L. redshift by the eGaussian model; $\emph{Column 8:}$ reproduction distance of image from the observed location in the eGaussian model; $\emph{Column 9:}$ comments.}
%\end{center}
\end{table*}

\subsection{Results and comparison of the two models}\label{comp}

In total, we found 70 multiple images and candidates of 23 background sources (Table \ref{multTable}, Fig. \ref{multipleimages}). All images \emph{not} marked as \emph{candidates} were used as constraints for the model: 34 images of 13 sources. For seven of these the redshifts were left to be optimized by the models. Two of the multiply-imaged sources seem to be high redshift ``dropouts'': system 6 at $z\sim6.5$, and candidate system 20 at $z\sim5$, although the redshift constraints on the latter are poor. We leave the detailed examination of these two objects for future work.

Both models yield very similar critical curves, except for a small region of a few arcseconds discrepancy where there are only systems that were not used as constraints (or their redshifts left free). In addition, the mass profiles (centered on the midpoint between the first and second brightest members, RA$_{J2000}$=04:16:08.38, DEC$_{J2000}$=-24:04:20.80) of the two models are almost identical throughout the range where multiple images are observed ($r<1\arcmin$), see Fig. \ref{profile}. The two models are also in excellent agreement with independent, color-color selected background galaxies \citep{Medezinski2010} WL measurements from $BR_{\rm c}z'$ Subaru data (Fig. \ref{profile}), using the Bayesian method of \citet{Umetsu2011b,Umetsu2011} that combines tangential-distortion and magnification-bias measurements in a model-independent manner, effectively breaking the mass-sheet degeneracy. The WL analysis pipeline is described in \citet{Umetsu2012}.

The image-plane reproduction rms($\chi^{2}$) is 1.89\arcsec(56.67) and 1.37\arcsec(29.67) for the eGaussian and eNFW models, respectively. The rms increases slightly to 2.40\arcsec and 2.11\arcsec, respectively, when including all images and candidates and not only those used as constraints. For the $\chi^{2}$ we used a positional error of $\sigma_{pos}=1.4\arcsec$, which was found to be a reasonable value accounting for large-scale structure and matter along the line-of-sight \citep[see][and references therein]{Zitrin2012CLASH1206}. The multiple-images input comprises 35 constraints, where the number of DOF is 26 and 22 for the eGaussian and eNFW models, respectively, yielding correspondingly $\chi^{2}/DOF$=2.18 and $\chi^{2}/DOF$=1.35. Both models show a critical area (A) with an effective Einstein radius of $\theta_{e}=\sqrt\frac{A}{\pi}\simeq26\pm2\arcsec$ ($z_{s}=1.896$), enclosing $M(<\theta_{e})=1.25\pm0.09 \times10^{14}M_{\odot}$. Comparing the two methods, the eNFW model has a somewhat better fit to the data - and higher flexibility to match it. The eGaussian smoothed model fit is somewhat worse, following more rigourously the light-traces-mass assumption, but the fact that it physically matches multiple images \emph{a priori} (without needing to accurately constrain the fit first), is remarkable evidence for its credibility.

The number density of multiple images and candidates uncovered, i.e. number per the given critical area (70 images of 23 sources over $\simeq0.6~\square\arcmin$ for $z_{s}\sim2$), is a few times higher than other prominent known lensing clusters (for similar background redshifts). A1689 ($z=0.19$) for example, one of the largest lenses known, shows 135 images of 42 sources, over a critical area of $\simeq1.8~\square\arcmin$ \citep{Coe2010}. A1703 ($z=0.28$) has more than 50 multiple images of 17 sources known, and a critical area of $\simeq0.8~\square\arcmin$ \citep[][]{Richard2009_A1703}. MACS J0717.5+3745 ($z=0.55$), the largest gravitational lens \citep{Zitrin2009_macs0717}, has about 60 multiple images known from 18 sources \citep[see also][]{Limousin2012_M0717}, and a critical area of $\simeq2.64~\square\arcmin$. %This high number density of multiple images in M0416 is also related to high magnification power in the center - which is indeed evident by some counter-images far from the critical area that are too faint and small to be unambiguously identified, implying as expected a much higher magnification of images straddling the critical curves.

%An important point to make is that although we are being conservative designating ambiguous cases as \emph{candidates}, the images listed here exist in the data, and according to their position and $z_{phot}$, should be multiply lensed. It can also be assumed, that with further dedicated effort more multiply-imaged sources will be found in M0416. Therefore, the number of multiple images is not likely to be lower than the number we quote here.

\section{Expected effect of ellipticity and merger}
We now wish to test the effects of ellipticity and merger on the observed number of multiple images.

We start by producing a fiducial eNFW model at $z=0.42$, with point sources planted behind it ($z_{s}=2$), every 45 kpc on a grid. We then increase the ellipticity ($e=(MJA-MNA)/(MJA+MNA)$) from 0.0 to 0.8, and count the number of multiple images generated. Note that throughout we always count \emph{all} multiple images formed. For each configuration we also measure the critical area for normalization, so the effect of ellipticity can be extracted per given critical area.

\begin{figure}
 \begin{center}
   \includegraphics[width=90mm]{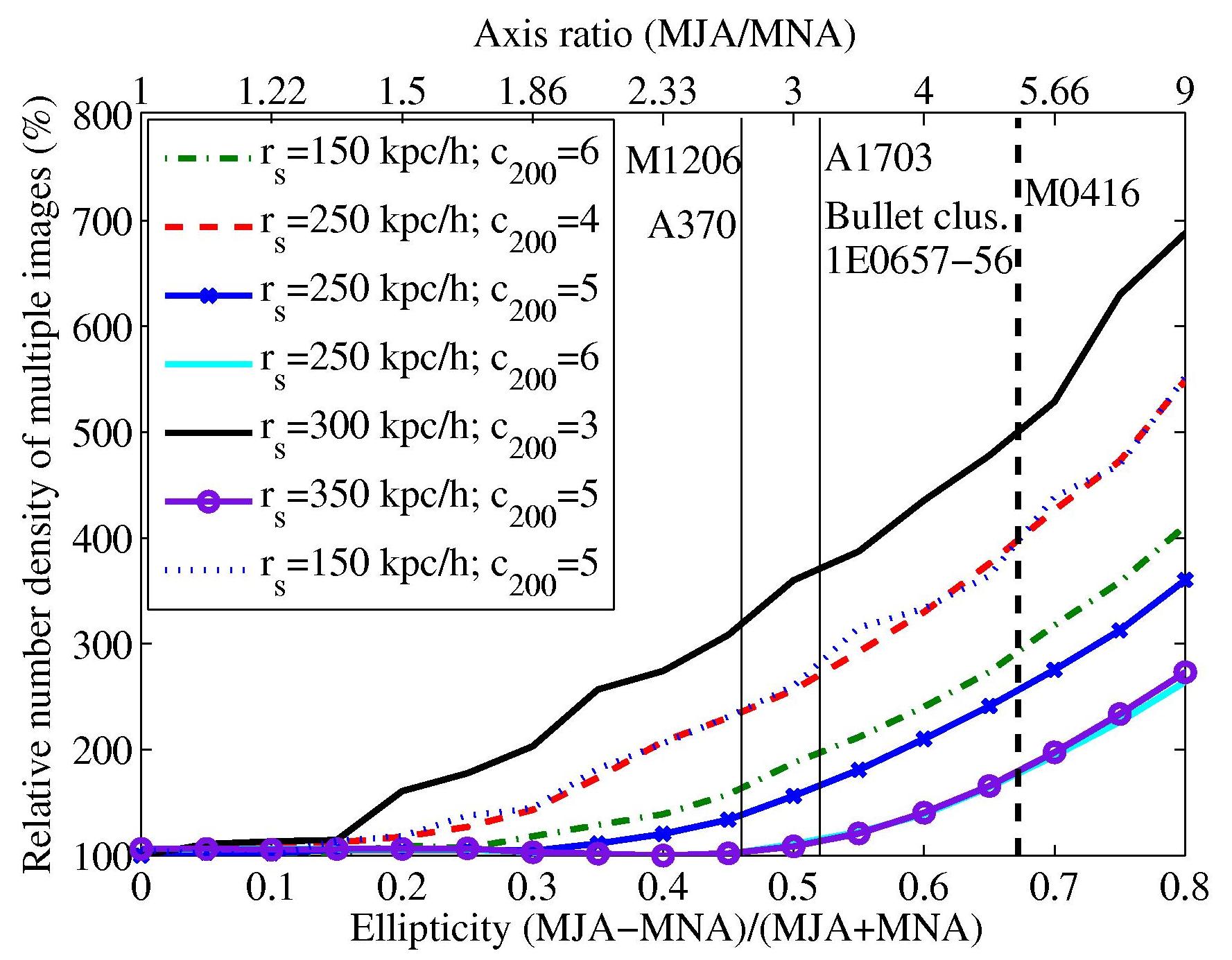}
 \end{center}
\caption{Number density of multiple images as a function of lens ellipticity (normalized to the circular case), for different concentration and scale-radius parameters. In all probed cases, higher ellipticity boosts the lensing efficiency, generating more multiple images \emph{per critical area}. We mark on the figure the measured effective ellipticity of M0416 and other (less) elongated clusters. Choosing those combinations of $c_{200}$ and $r_{s}$ that yield comparable critical area to that of M0416, we obtain that the number density of multiple images observed in M0416 is $\sim2.5\times$ higher due its elongation.}
\label{increaseEll}
\end{figure}

\begin{figure}
 \begin{center}
   \includegraphics[width=92mm]{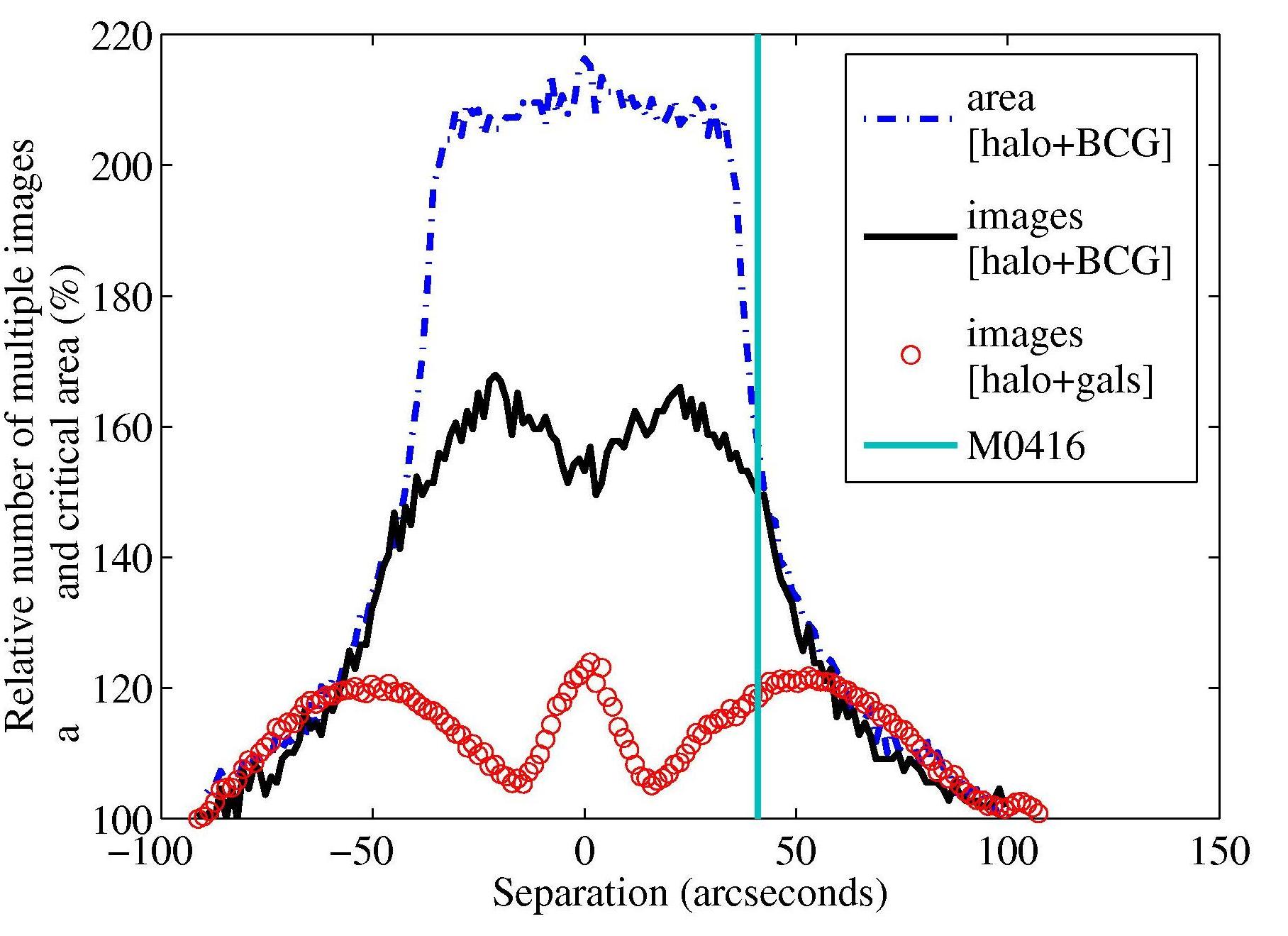}
 \end{center}
\caption{Critical area (\emph{Blue dash-dotted line}) and number of multiple images versus the displacement between the two halos, normalized to a large-separation case. The \emph{Black curve} shows the behavior when accounting only for the DM halos and BCGs, while the \emph{Red circle curve} shows the same scenario but with more massive halos to account for the missing galaxies mass. In either case, it is clear that the merger contributes a boost of $\sim20-60\%$ to the number of multiple images observed.}
\label{increaseMerge}
\end{figure}

In Fig. \ref{increaseEll} we plot the resulting increase in the number of multiple images with lens ellipticity, per critical area, for various combinations of NFW parameters. A clear correlation is observed, so larger numbers of multiple images are generated by higher ellipticities (but the amplitude may vary with masses and distances). To assess the effect in M0416 (for a fixed axis ratio of $\simeq5.1$), we choose different combinations of $c_{200}$ and $r_{s}$ that yield comparable critical area to that of M0416. The observed elongation results in a $\sim2.5\times$ increase in the number of multiple images compared with the circular case, for the given critical area. Therefore, the observed elongation explains why M0416 has a few times higher multiple-image number density than other typical lensing clusters (in comparable HST imaging). In reality, the increase in the number density of multiple images with ellipticity is of course, a more complex function and the exact number depends also on the luminosity function and observational depth, and on the mass shape parameters. However here our goal was to merely show that such a correlation exists and assess its order of magnitude.

To test the effect of merger in M0416, we simulate two eNFW halos with similar parameters as in the resulting mass model including also the BCGs in their centers, approaching each other on the line connecting them (Fig. \ref{increaseMerge}). The observed separation between the two halos entails a $\sim20-60\%(120)\%$ increase in the total number of multiple-images(critical area), compared to the ``far-away'' initial position where each halo comprised its own (unmerged) critical curve. The exact amount is dependent on the true mass of each halo, including the true contribution of the galaxies (Fig. \ref{increaseMerge}). Note that as the two halos continue approaching each other, although the total number of multiple images may increase, the number of multiple images \emph{per critical area}, will decrease.

%In addition, note that since along the major axis there is a third bright member south of the second BCG - we acknowledge the possibility that its halo could additionally add to the effect - but neglect it here since in our SL fit there was no need to add a third halo at its location.

To assess how extreme is the elongation of the critical curves, we compare to Multidark/MUSIC 2 numerical simulations \citep{Sembolini2012Music}. We used the 80 clusters at $z\simeq0.42$ above the completeness mass limit $M_{vir}=6 \times10^{14} h^{-1}M_{\odot}$ in a volume of 1$h^{-3}$Gpc$^{3}$. Each cluster was projected along 100 lines of sight. We searched for those projections where the critical lines both exceed $50\arcsec$ in at least one direction, and exhibit high elongations (axis ratio $>5$). These conditions are satisfied in $\sim4\%$ of the 8000 lens planes.

\section{Summary}
We presented a SL study of M0416, in which we uncovered 70 multiple images and candidates of 23 background sources. We constructed mass models using two independent methods, both yield similar critical curves and mass profiles, in agreement also with independent larger-scale WL analysis. Compared to other well-known lensing clusters, M0416 exhibits a high number of multiple images for its critical area ($\simeq0.6~\square\arcmin$).

We simulated the effects of ellipticity and merger on the lensing efficiency, and showed that (a) the \emph{number density} of multiple images increases with ellipticity (the source-plane caustics get bigger, generating more multiple images \emph{for the same critical area}), and (b) the critical area, and correspondingly \emph{total number} of multiple images, increase with lower separations between two merging clumps, peaking when the two halos are either on top of each other or a few dozen arcseconds away, depending on their mass and shape.

We conclude that the observed critical area size can be attributed to the merger, which boosts the mass in the center. For this given critical area, the high multiple-image number density can be explained by the observed elongation, which boosts the lensing efficiency by $\sim2.5\times$. Background cosmic variance, estimated typically at a $\sim20\%$ level \citep{Somerville2004CosmicVariance} or double for $z>5$ \citep{TrentiStiavelli2008}, is likely to play only a small role in the increased number of multiple images, which are spread over a wide redshift range.

This cluster shows once more that there exists a class of prominent lenses at redshifts around $z\sim0.4-0.5$ (and higher), probably due to their merging state and thus level of substructure and ellipticity which as we have shown, boost the SL properties.

\section*{acknowledgments}

We thank the anonymous reviewer of this work for very valuable comments. AZ is supported by contract research ``"Internationale Spitzenforschung II/2-6'' of the Baden W\"urttemberg Stiftung. ACS was developed under NASA contract NAS 5-32865. Results are based on observations made with
the NASA/ESA Hubble Space Telescope, obtained from the data archive at
the Space Telescope Science Institute. STScI is operated by the
Association of Universities for Research in Astronomy, Inc. under NASA
contract NAS 5-26555. This work is based in part on data collected at the Subaru Telescope,
which is operated by the National Astronomical Society of
Japan. PR acknowledges partial support by the DFG cluster of excellence Origin and Structure of the Universe. KU acknowledges partial support from the National Science Council of Taiwan grant NSC100-2112-M-001-008-MY3 and from the Academia Sinica Career Development Award. The work of LAM was carried out at Jet Propulsion Laboratory, California Institute of Technology, under a contract with NASA.

\bibliographystyle{apj}
\bibliography{outDan2}

\begin{thebibliography}{50}
\expandafter\ifx\csname natexlab\endcsname\relax\def\natexlab#1{#1}\fi

\bibitem[{{Bartelmann}(2010)}]{Bartelmann2010reviewB}
{Bartelmann}, M. 2010, Classical and Quantum Gravity, 27, 233001

\bibitem[{{Ben{\'{\i}}tez} {et~al.}(2004){Ben{\'{\i}}tez}, {Ford}, {Bouwens},
  {Menanteau}, {Blakeslee}, {Gronwall}, {Illingworth}, {Meurer}, {Broadhurst},
  {Clampin}, {et~al.}}]{Benitez2004}
{Ben{\'{\i}}tez}, N., {Ford}, H., {Bouwens}, R., {et~al.} 2004, \apjs, 150, 1

\bibitem[{{Bertin} \& {Arnouts}(1996)}]{BertinArnouts1996Sextractor}
{Bertin}, E., \& {Arnouts}, S. 1996, \aaps, 117, 393

\bibitem[{{Brada{\v c}} {et~al.}(2006){Brada{\v c}}, {Clowe}, {Gonzalez},
  {Marshall}, {Forman}, {Jones}, {Markevitch}, {Randall}, {Schrabback}, \&
  {Zaritsky}}]{Bradac2006Bullet}
{Brada{\v c}}, M., {Clowe}, D., {Gonzalez}, A.~H., {et~al.} 2006, \apj, 652,
  937

\bibitem[{{Broadhurst} {et~al.}(2008){Broadhurst}, {Umetsu}, {Medezinski},
  {Oguri}, \& {Rephaeli}}]{Broadhurst2008}
{Broadhurst}, T., {Umetsu}, K., {Medezinski}, E., {Oguri}, M., \& {Rephaeli},
  Y. 2008, \apjl, 685, L9

\bibitem[{{Broadhurst} {et~al.}(2005){Broadhurst}, {Ben{\'{\i}}tez}, {Coe},
  {Sharon}, {Zekser}, {White}, {Ford}, {Bouwens}, {Blakeslee}, {Clampin},
  {et~al.}}]{Broadhurst2005a}
{Broadhurst}, T., {Ben{\'{\i}}tez}, N., {Coe}, D., {et~al.} 2005, \apj, 621, 53

\bibitem[{{Broadhurst} \& {Barkana}(2008)}]{BroadhurstBarkana2008}
{Broadhurst}, T.~J., \& {Barkana}, R. 2008, \mnras, 390, 1647

\bibitem[{{Christensen} {et~al.}(2012){Christensen}, {Richard}, {Hjorth},
  {Milvang-Jensen}, {Laursen}, {Limousin}, {Dessauges-Zavadsky}, {Grillo}, \&
  {Ebeling}}]{Christensen2012Specs}
{Christensen}, L., {Richard}, J., {Hjorth}, J., {et~al.} 2012, arXiv, 1209.0767

\bibitem[{{Coe} {et~al.}(2010){Coe}, {Ben{\'{\i}}tez}, {Broadhurst}, \&
  {Moustakas}}]{Coe2010}
{Coe}, D., {Ben{\'{\i}}tez}, N., {Broadhurst}, T., \& {Moustakas}, L.~A. 2010,
  \apj, 723, 1678

\bibitem[{{Coe} {et~al.}(2006){Coe}, {Ben{\'{\i}}tez}, {S{\'a}nchez}, {Jee},
  {Bouwens}, \& {Ford}}]{Coe2006}
{Coe}, D., {Ben{\'{\i}}tez}, N., {S{\'a}nchez}, S.~F., {et~al.} 2006, \aj, 132,
  926

\bibitem[{{Coe} {et~al.}(2012){Coe}, {Zitrin}, {Carrasco}, {Shu}, {Zheng},
  {Postman}, {Bradley}, {Koekemoer}, {Bouwens}, {Broadhurst}, {Monna}, {Host},
  {Moustakas}, {Ford}, {Moustakas}, {van der Wel}, {Donahue}, {Rodney},
  {Benitez}, {Jouvel}, {Seitz}, {Kelson}, \& {Rosati}}]{Coe2012highz}
{Coe}, D., {Zitrin}, A., {Carrasco}, M., {et~al.} 2012, arXiv, 1211.3663

\bibitem[{{Comerford} \& {Natarajan}(2007)}]{ComerfordNatarajan2007CMrelation}
{Comerford}, J.~M., \& {Natarajan}, P. 2007, \mnras, 379, 190

\bibitem[{{Ebeling} {et~al.}(2007){Ebeling}, {Barrett}, {Donovan}, {Ma},
  {Edge}, \& {van Speybroeck}}]{EbelingMacs12_2007}
{Ebeling}, H., {Barrett}, E., {Donovan}, D., {et~al.} 2007, \apjl, 661, L33

\bibitem[{{Ebeling} {et~al.}(2010){Ebeling}, {Edge}, {Mantz}, {Barrett},
  {Henry}, {Ma}, \& {van Speybroeck}}]{Ebeling2010FinalMACS}
{Ebeling}, H., {Edge}, A.~C., {Mantz}, A., {et~al.} 2010, \mnras, 407, 83

\bibitem[{{Halkola} {et~al.}(2006){Halkola}, {Seitz}, \&
  {Pannella}}]{Halkola2006}
{Halkola}, A., {Seitz}, S., \& {Pannella}, M. 2006, \mnras, 372, 1425

\bibitem[{{Hennawi} {et~al.}(2007){Hennawi}, {Dalal}, {Bode}, \&
  {Ostriker}}]{Hennawi2007}
{Hennawi}, J.~F., {Dalal}, N., {Bode}, P., \& {Ostriker}, J.~P. 2007, \apj,
  654, 714

\bibitem[{{Jauzac} {et~al.}(2012){Jauzac}, {Jullo}, {Kneib}, {Ebeling},
  {Leauthaud}, {Ma}, {Limousin}, {Massey}, \& {Richard}}]{Jauzac2012WL0717}
{Jauzac}, M., {Jullo}, E., {Kneib}, J.-P., {et~al.} 2012, \mnras, 426, 3369

\bibitem[{{Jullo} {et~al.}(2007){Jullo}, {Kneib}, {Limousin},
  {El{\'{\i}}asd{\'o}ttir}, {Marshall}, \& {Verdugo}}]{Jullo2007Lenstool}
{Jullo}, E., {Kneib}, J.-P., {Limousin}, M., {et~al.} 2007, New Journal of
  Physics, 9, 447

\bibitem[{{Kassiola} \& {Kovner}(1993)}]{KassiolaKovner1993}
{Kassiola}, A., \& {Kovner}, I. 1993, \apj, 417, 450

\bibitem[{{Kneib} \& {Natarajan}(2011)}]{Kneib2011review}
{Kneib}, J.-P., \& {Natarajan}, P. 2011, \aapr, 19, 47

\bibitem[{{Limousin} {et~al.}(2012){Limousin}, {Ebeling}, {Richard},
  {Swinbank}, {Smith}, {Jauzac}, {Rodionov}, {Ma}, {Smail}, {Edge}, {Jullo}, \&
  {Kneib}}]{Limousin2012_M0717}
{Limousin}, M., {Ebeling}, H., {Richard}, J., {et~al.} 2012, \aap, 544, A71

\bibitem[{{Mann} \& {Ebeling}(2012)}]{MannEbeling2012evolution}
{Mann}, A.~W., \& {Ebeling}, H. 2012, \mnras, 420, 2120

\bibitem[{{Medezinski} {et~al.}(2010){Medezinski}, {Broadhurst}, {Umetsu},
  {Oguri}, {Rephaeli}, \& {Ben{\'{\i}}tez}}]{Medezinski2010}
{Medezinski}, E., {Broadhurst}, T., {Umetsu}, K., {et~al.} 2010, \mnras, 405,
  257

\bibitem[{{Meneghetti} {et~al.}(2007){Meneghetti}, {Argazzi}, {Pace},
  {Moscardini}, {Dolag}, {Bartelmann}, {Li}, \& {Oguri}}]{meneghetti2007}
{Meneghetti}, M., {Argazzi}, R., {Pace}, F., {et~al.} 2007, \aap, 461, 25

\bibitem[{{Meneghetti} {et~al.}(2003){Meneghetti}, {Bartelmann}, \&
  {Moscardini}}]{Meneghetti2003}
{Meneghetti}, M., {Bartelmann}, M., \& {Moscardini}, L. 2003, \mnras, 346, 67

\bibitem[{{Merten} {et~al.}(2011){Merten}, {Coe}, {Dupke}, {Massey}, {Zitrin},
  {Cypriano}, {Okabe}, {Frye}, {Braglia}, {Jim{\'e}nez-Teja}, {Ben{\'{\i}}tez},
  {Broadhurst}, {Rhodes}, {Meneghetti}, {Moustakas}, {Sodr{\'e}}, {Krick}, \&
  {Bregman}}]{Merten2011}
{Merten}, J., {Coe}, D., {Dupke}, R., {et~al.} 2011, \mnras, 417, 333

\bibitem[{{Oguri} {et~al.}(2012{\natexlab{a}}){Oguri}, {Bayliss}, {Dahle},
  {Sharon}, {Gladders}, {Natarajan}, {Hennawi}, \&
  {Koester}}]{Oguri201238clusters}
{Oguri}, M., {Bayliss}, M.~B., {Dahle}, H., {et~al.} 2012{\natexlab{a}},
  \mnras, 420, 3213

\bibitem[{{Oguri} {et~al.}(2012{\natexlab{b}}){Oguri}, {Schrabback}, {Jullo},
  {Ota}, {Kochanek}, {Dai}, {Ofek}, {Richards}, {Blandford}, {Falco}, \&
  {Fohlmeister}}]{Oguri2012SL}
{Oguri}, M., {Schrabback}, T., {Jullo}, E., {et~al.} 2012{\natexlab{b}}, arXiv,
  1209.0458

\bibitem[{{Okabe} {et~al.}(2010){Okabe}, {Takada}, {Umetsu}, {Futamase}, \&
  {Smith}}]{Okabe2010}
{Okabe}, N., {Takada}, M., {Umetsu}, K., {Futamase}, T., \& {Smith}, G.~P.
  2010, \pasj, 62, 811

\bibitem[{{Postman} {et~al.}(2012){Postman}, {Coe}, {Ben{\'{\i}}tez},
  {Bradley}, {Broadhurst}, {Donahue}, {Ford}, {Graur}, {Graves}, {Jouvel},
  {Koekemoer}, {Lemze}, {Medezinski}, {Molino}, {Moustakas}, {Ogaz}, {Riess},
  {Rodney}, {Rosati}, {Umetsu}, {Zheng}, {Zitrin}, {Bartelmann}, {Bouwens},
  {Czakon}, {Golwala}, {Host}, {Infante}, {Jha}, {Jimenez-Teja}, {Kelson},
  {Lahav}, {Lazkoz}, {Maoz}, {McCully}, {Melchior}, {Meneghetti}, {Merten},
  {Moustakas}, {Nonino}, {Patel}, {Reg{\"o}s}, {Sayers}, {Seitz}, \& {Van der
  Wel}}]{PostmanCLASHoverview}
{Postman}, M., {Coe}, D., {Ben{\'{\i}}tez}, N., {et~al.} 2012, \apjs, 199, 25

\bibitem[{{Redlich} {et~al.}(2012){Redlich}, {Bartelmann}, {Waizmann}, \&
  {Fedeli}}]{Redlich2012MergerRE}
{Redlich}, M., {Bartelmann}, M., {Waizmann}, J.-C., \& {Fedeli}, C. 2012,
  arXiv, 1205.6906

\bibitem[{{Richard} {et~al.}(2010){Richard}, {Kneib}, {Limousin}, {Edge}, \&
  {Jullo}}]{Richard2010A370}
{Richard}, J., {Kneib}, J.-P., {Limousin}, M., {Edge}, A., \& {Jullo}, E. 2010,
  \mnras, 402, L44

\bibitem[{{Richard} {et~al.}(2009){Richard}, {Pei}, {Limousin}, {Jullo}, \&
  {Kneib}}]{Richard2009_A1703}
{Richard}, J., {Pei}, L., {Limousin}, M., {Jullo}, E., \& {Kneib}, J.~P. 2009,
  \aap, 498, 37

\bibitem[{{Sadeh} \& {Rephaeli}(2008)}]{SadehRephaeli2008}
{Sadeh}, S., \& {Rephaeli}, Y. 2008, \mnras, 388, 1759

\bibitem[{{Sembolini} {et~al.}(2012){Sembolini}, {Yepes}, {De Petris},
  {Gottloeber}, {Lamagna}, \& {Comis}}]{Sembolini2012Music}
{Sembolini}, F., {Yepes}, G., {De Petris}, M., {et~al.} 2012, arXiv, 1207.4438

\bibitem[{{Sereno} {et~al.}(2010){Sereno}, {Jetzer}, \& {Lubini}}]{Sereno2010}
{Sereno}, M., {Jetzer}, P., \& {Lubini}, M. 2010, \mnras, 403, 2077

\bibitem[{{Sereno} \& {Zitrin}(2012)}]{SerenoZitrin2012}
{Sereno}, M., \& {Zitrin}, A. 2012, \mnras, 419, 3280

\bibitem[{{Somerville} {et~al.}(2004){Somerville}, {Lee}, {Ferguson},
  {Gardner}, {Moustakas}, \& {Giavalisco}}]{Somerville2004CosmicVariance}
{Somerville}, R.~S., {Lee}, K., {Ferguson}, H.~C., {et~al.} 2004, \apjl, 600,
  L171

\bibitem[{{Torri} {et~al.}(2004){Torri}, {Meneghetti}, {Bartelmann},
  {Moscardini}, {Rasia}, \& {Tormen}}]{Torri2004}
{Torri}, E., {Meneghetti}, M., {Bartelmann}, M., {et~al.} 2004, \mnras, 349,
  476

\bibitem[{{Trenti} \& {Stiavelli}(2008)}]{TrentiStiavelli2008}
{Trenti}, M., \& {Stiavelli}, M. 2008, \apj, 676, 767

\bibitem[{{Umetsu} {et~al.}(2011{\natexlab{a}}){Umetsu}, {Broadhurst},
  {Zitrin}, {Medezinski}, {Coe}, \& {Postman}}]{Umetsu2011b}
{Umetsu}, K., {Broadhurst}, T., {Zitrin}, A., {et~al.} 2011{\natexlab{a}},
  \apj, 738, 41

\bibitem[{{Umetsu} {et~al.}(2011{\natexlab{b}}){Umetsu}, {Broadhurst},
  {Zitrin}, {Medezinski}, \& {Hsu}}]{Umetsu2011}
{Umetsu}, K., {Broadhurst}, T., {Zitrin}, A., {Medezinski}, E., \& {Hsu}, L.
  2011{\natexlab{b}}, \apj, 729, 127

\bibitem[{{Umetsu} {et~al.}(2012){Umetsu}, {Medezinski}, {Nonino}, {Merten},
  {Zitrin}, {Molino}, {Grillo}, {Carrasco}, {Donahue}, {Mahdavi}, {Coe},
  {Postman}, {Koekemoer}, {Czakon}, {Sayers}, {Mroczkowski}, {Golwala}, {Koch},
  {Lin}, {Molnar}, {Rosati}, {Balestra}, {Mercurio}, {Scodeggio}, {Biviano},
  {Anguita}, {Infante}, {Seidel}, {Sendra}, {Jouvel}, {Host}, {Lemze},
  {Broadhurst}, {Meneghetti}, {Moustakas}, {Bartelmann}, {Ben{\'{\i}}tez},
  {Bouwens}, {Bradley}, {Ford}, {Jim{\'e}nez-Teja}, {Kelson}, {Lahav},
  {Melchior}, {Moustakas}, {Ogaz}, {Seitz}, \& {Zheng}}]{Umetsu2012}
{Umetsu}, K., {Medezinski}, E., {Nonino}, M., {et~al.} 2012, \apj, 755, 56

\bibitem[{{Zheng} {et~al.}(2012){Zheng}, {Postman}, {Zitrin}, {Moustakas},
  {Shu}, {Jouvel}, {H{\o}st}, {Molino}, {Bradley}, {Coe}, {Moustakas},
  {Carrasco}, {Ford}, {Ben{\'{\i}}tez}, {Lauer}, {Seitz}, {Bouwens},
  {Koekemoer}, {Medezinski}, {Bartelmann}, {Broadhurst}, {Donahue}, {Grillo},
  {Infante}, {Jha}, {Kelson}, {Lahav}, {Lemze}, {Melchior}, {Meneghetti},
  {Merten}, {Nonino}, {Ogaz}, {Rosati}, {Umetsu}, \& {van der
  Wel}}]{Zheng2012NaturZ}
{Zheng}, W., {Postman}, M., {Zitrin}, A., {et~al.} 2012, \nat, 489, 406

\bibitem[{{Zitrin} \& {Broadhurst}(2009)}]{Zitrin2009_macs11495}
{Zitrin}, A., \& {Broadhurst}, T. 2009, \apjl, 703, L132

\bibitem[{{Zitrin} {et~al.}(2011){Zitrin}, {Broadhurst}, {Barkana}, {Rephaeli},
  \& {Ben{\'{\i}}tez}}]{Zitrin2011_12macsclusters}
{Zitrin}, A., {Broadhurst}, T., {Barkana}, R., {Rephaeli}, Y., \&
  {Ben{\'{\i}}tez}, N. 2011, \mnras, 410, 1939

\bibitem[{{Zitrin} {et~al.}(2009{\natexlab{a}}){Zitrin}, {Broadhurst},
  {Rephaeli}, \& {Sadeh}}]{Zitrin2009_macs0717}
{Zitrin}, A., {Broadhurst}, T., {Rephaeli}, Y., \& {Sadeh}, S.
  2009{\natexlab{a}}, \apjl, 707, L102

\bibitem[{{Zitrin} {et~al.}(2009{\natexlab{b}}){Zitrin}, {Broadhurst},
  {Umetsu}, {Coe}, {Ben{\'{\i}}tez}, {Ascaso}, {Bradley}, {Ford}, {Jee},
  {Medezinski}, {Rephaeli}, \& {Zheng}}]{Zitrin2009_cl0024}
{Zitrin}, A., {Broadhurst}, T., {Umetsu}, K., {et~al.} 2009{\natexlab{b}},
  \mnras, 396, 1985

\bibitem[{{Zitrin} {et~al.}(2012{\natexlab{a}}){Zitrin}, {Moustakas},
  {Bradley}, {Coe}, {Moustakas}, {Postman}, {Shu}, {Zheng}, {Ben{\'{\i}}tez},
  {Bouwens}, {Broadhurst}, {Ford}, {Host}, {Jouvel}, {Koekemoer}, {Meneghetti},
  {Rosati}, {Donahue}, {Grillo}, {Kelson}, {Lemze}, {Medezinski}, {Molino},
  {Nonino}, \& {Ogaz}}]{Zitrin2012CLASH0329}
{Zitrin}, A., {Moustakas}, J., {Bradley}, L., {et~al.} 2012{\natexlab{a}},
  \apjl, 747, L9

\bibitem[{{Zitrin} {et~al.}(2012{\natexlab{b}}){Zitrin}, {Rosati}, {Nonino},
  {Grillo}, {Postman}, {Coe}, {Seitz}, {Eichner}, {Broadhurst}, {Jouvel},
  {Balestra}, {Mercurio}, {Scodeggio}, {Ben{\'{\i}}tez}, {Bradley}, {Ford},
  {Host}, {Jimenez-Teja}, {Koekemoer}, {Zheng}, {Bartelmann}, {Bouwens},
  {Czoske}, {Donahue}, {Graur}, {Graves}, {Infante}, {Jha}, {Kelson}, {Lahav},
  {Lazkoz}, {Lemze}, {Lombardi}, {Maoz}, {McCully}, {Medezinski}, {Melchior},
  {Meneghetti}, {Merten}, {Molino}, {Moustakas}, {Ogaz}, {Patel}, {Regoes},
  {Riess}, {Rodney}, {Umetsu}, \& {Van der Wel}}]{Zitrin2012CLASH1206}
{Zitrin}, A., {Rosati}, P., {Nonino}, M., {et~al.} 2012{\natexlab{b}}, \apj,
  749, 97

\end{thebibliography}

\end{document}